\documentclass{article}

\usepackage{arxiv}

\usepackage[utf8]{inputenc} 
\usepackage[T1]{fontenc}    
\usepackage{hyperref}       
\usepackage{url}            
\usepackage{booktabs}       
\usepackage{amsfonts}       
\usepackage{nicefrac}       
\usepackage{microtype}      
\usepackage{graphicx}
\usepackage[numbers]{natbib}
\usepackage{doi}
\usepackage{enumitem}
\usepackage{makecell}
\usepackage{titlesec}
\usepackage{amsmath}
\usepackage[T2A]{fontenc}

\setlist{leftmargin=3.0mm}

\titlespacing\subsection{0pt}{12pt plus 4pt minus 2pt}{0pt plus 4pt minus 2pt}
\title{Artificial bees collect diverse conformers of small organic molecules}

\author{
    {Anastasiia Smirnova} \\
	Department of Chemistry\\
	Lomonosov Moscow State University\\
	Moscow 119991, Russia\\
	\And
    {Maxim Yablonskiy} \\
	Department of Chemistry\\
	Lomonosov Moscow State University\\
	Moscow 119991, Russia\\
	\And
    {Ekaterina Marchenko} \\
	Department of Geology\\
	Lomonosov Moscow State University\\
	Moscow 119991, Russia\\
	\And
    \href{https://orcid.org/0000-0001-6117-5662}{\includegraphics[scale=0.07]{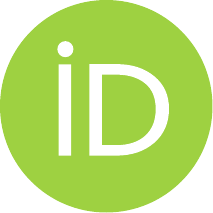}\hspace{1mm}Vadim Korolev} \\
	Department of Chemistry\\
	Lomonosov Moscow State University\\
	Moscow 119991, Russia\\
	\And
    \href{http://orcid.org/0000-0003-2020-1953}{\includegraphics[scale=0.07]{orcid.pdf}\hspace{1mm}Nikolai Andreadi} \\
	Department of Chemistry\\
	Lomonosov Moscow State University\\
	Moscow 119991, Russia\\
	\And
	\href{http://orcid.org/0000-0001-8891-6862}{\includegraphics[scale=0.07]{orcid.pdf}\hspace{1mm}Artem Mitrofanov}\thanks{\textit{Email address}: \texttt{mitrofjr@radio.chem.msu.ru}} \\
	Department of Chemistry\\
	Lomonosov Moscow State University\\
	Moscow 119991, Russia\\
}

\renewcommand{\shorttitle}{Artificial bees collect diverse conformers of small organic molecules}

\begin{document}
\maketitle

\begin{abstract}
The conformational mobility of organic molecules defined as a variability of practically accessible conformers plays a critical role in determining electronic, chemical, and physical properties within computational methods. At the same time, there is a challenge of identifying compact set of global and local conformers for the comprehensive description of potential energy surface. Here we apply nature-inspired algorithms to resolve this issue. Among all the considered algorithms, the artificial bee colony optimizer exhibits the highest performance in discovering conformers detected both in gas and condensed phases. We hope that our approach enables researchers to make a next step in organic crystal packing studies.
\end{abstract}

\section{Introduction}
\label{sec:introduction}
Conformational analysis is an important part of chemical design, allowing to evaluate the contribution of the flexibility of small organic molecules to the thermodynamics and kinetics of many processes, ranging from complex formation\cite{mitrofanov2021iii,rankin2013structural} to molecular packing in organic crystals\cite{townes2013microwave}. In addition, intelligent conformational subsampling serves as the basis for so-called 4D quantitative structure-property relationship (QSPR) approach, allowing information about the behavior of a small organic molecule in a gas or liquid phase to be conveyed to a machine learning algorithm\cite{broyden1967quasi}. From a physical point of view, the conformational analysis corresponds to the problem of searching for both global and local minima on the potential energy surface (PES) of a molecule. The issue is solved both by using a mixture of experimental methods (e.g., gas electron diffraction\cite{rankin2013structural} and microwave spectroscopy\cite{townes2013microwave}) and quantum chemical calculations, and by various computational approaches that are relied on the crystal structure data. However, such an approach is complicated by the computational cost of PES calculations and considering the rotational symmetry of molecules. Additionally, in a case of the solid-state data we faced with the discussion about the similarity of organic molecule conformations in a gas and condensed states\cite{thompson2014conformations}.

Currently, there are robust solutions of the problem of local optimization of molecular geometry\cite{broyden1967quasi,fletcher1970new,goldfarb1970family,shanno1970conditioning,bitzek2006structural}. Additionally, both physically inspired methods\cite{goedecker2004minima} and global optimization algorithms\cite{andreadi2022tree} offer alternative approaches for conducting a directed search towards achieving a global PES minimum. Methods for generating possible conformations have been also developed (from the distance geometry matrix based\cite{riniker2015better} to the driven by artificial intelligence (AI)\cite{wang2023small}). However, most of them relied on the crystal structure data; therefore, they may differ markedly from the gas-phase one. In addition, they often generated an excessive number of conformations, which could be used in 4D-QSPR or other relatively fast computational methods but are of little use in precise quantum chemical calculations.

Here we investigated the swarm intelligence methods to the conformational search issue. We needed to choose and modify the most efficient approach combining the directed search of the global minima with the exploration of the local ones using the satisfactory computational time (expressed in the number of calls for a single-point quantum chemical calculation). As a result, we built the physically consistent conformational analysis pipeline based on artificial bee colony algorithm and gas-phase semiempirical quantum-chemical calculations and we tested it on the manually curated experimental gas-phase data. We examined the previously collected molecular database and code for benchmarking of global optimization algorithms to choose the most appropriate approach to be placed in the basement of the conformational analysis method. We also discussed a possibility of its extension to crystal packaging as a first stage of solving crystal structure prediction problem.

\section{Computational details}
Here we investigated a series of the nature-inspired algorithms to solve the global optimization problem. There were 16 approaches: Firefly\cite{yang2010firefly}, Particle Swarm\cite{eberhart1995new}, Bees\cite{pham2006bees}, Artificial Bees Colony\cite{karaboga2005idea}, Bat\cite{gonzalez2008nature}, Flower Pollination\cite{yang2012flower}, Camel\cite{ibrahim2016novel}, Harris Hawks\cite{heidari2019harris}, Cat Swarm\cite{chu2006cat}, Grey Wolf\cite{mirjalili2014grey}, Moth Flame\cite{shehab2020moth}, Glow Worm Swarm\cite{krishnanand2009glowworm}, Cuckoo\cite{gandomi2013cuckoo}, Fish School\cite{bastos2008novel}, Harmony\cite{geem2001new}, and Monkey King Evolution\cite{meng2016monkey}. The experimental reference geometry data was taken from the Database of small organic molecules in the global minimum conformation\cite{andreadi2022tree} (gminorgDB). The RDKit library and NiaPy package\cite{vrbanvcivc2018niapy} were used for molecular data processing and conformation analysis within nature-inspired algorithms, respectively. We also used the OpenBabel package\cite{o2011open} to extract the conformations of molecules from the crystallographic data.

The existence of different conformations is based on the ability of molecule to rotate around single bonds connecting non-terminal atoms. At the same time, the formed structures are considered to belong to different local minima in the presence of a potential barrier on the reaction path from one geometry to another. A comparison of different conformations could be done by calculating room mean square deviation (RMSD) between atomic coordinates of two conformers. Since the RMSD calculation is carried out at every algorithm iteration, the operating time of this stage is critically important, especially when a set of conformations achieves several dozens. For this reason, we considered a set of dihedral angles instead of atomic coordinates as an inner similarity measure. The number of dihedral angles (significant Degrees of Freedom, DoFs) corresponded to the number of rotatable single bonds. Nevertheless, we used RMSD for atomic coordinates as a metric at the last stage owing to the omnipresence of this approach; the obtained results may be directly compared with the previous ones. The extended tight-binding quantum chemical method xTB\cite{bannwarth2021extended} (GFN2-xTB) was used to estimate the energy of conformations obtained during the operation of global optimization algorithms. The xTB offers a quick estimate of the energy of conformation while maintaining the accuracy of hybrid density functional functionals. We also validated the presence of potential barriers between the found conformations by the Nudged Elastic Band (NEB) method\cite{henkelman2000climbing} implemented in the xTB.

\section{Results and discussion}
\label{sec:results}
\subsection{Benchmarking of swarm intelligence methods}
At the first stage of comparing swarm intelligence methods, we make sure that they are able to determine the global minimum on the potential energy surface correctly. Here we use a deliberately large number of iterations to allow all approaches to successfully complete the optimization, considering the error of the energy calculation method. We utilize a dataset with information about global minima from the gminorg database as data for testing the algorithms.

All the considered methods achieve the equilibrium geometries with similar errors (RMSD/DoF values are less than 3.25 Å, including hydrogen positions; for more details see Fig. S1). Next, we consider the average number of steps that the methods take to reach the minimum of energy (Fig. 1). Several methods reach the minimum (up to the error of calculating the internal energy) noticeably faster, regardless of molecule size. It should be noted that at this step the local geometry optimization of intermediate conformers is not carried out due to its incompatibility with a substantial part of the swarm intelligence methods. As it follows from the data presented in Fig. 1, the best approaches are the Firefly, Particle Swarm, Bat, Flower Pollination, Harris Hawks, Cat Swarm, Grey Wolf, Harmony, and Cuckoo.

\begin{figure}[t!]
  \centering
  \includegraphics[width=9cm]{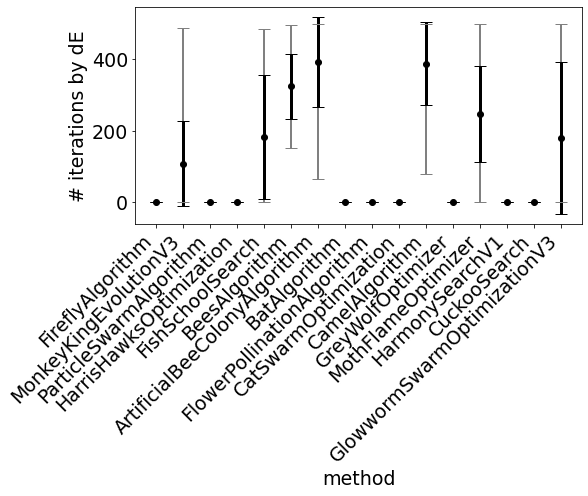}
  \caption{The average (point), 95\% confidence interval (black), and minimum and maximum (grey) number of iterations required to achieve a minimum on the potential energy surface of molecules.}
  \label{fig:fig1}
\end{figure}

Since the "one iteration" means a different number of calls for a quantum chemical calculation for different methods, we also compare the total running time of the algorithms for consistency. It should be noted that at this stage of the analysis we do not work on improving computational efficiency of algorithms, and all calculations are performed in a single-thread mode. Results are presented in Fig. 2. The fastest algorithm is the Harmonic search, which is a variation of the evolutionary approach. The following approaches after this method are the Grey Wolf Optimizer and the Artificial Bee Colony algorithm.

It is important to notice that we use a fast semi-empirical quantum chemical method for energy evaluation, and the calculation time per one geometry is on the order of tens of seconds. This time is only one or two orders of magnitude longer than the average time of the algorithm working for one iteration. If we consider the application of this method with other quantum chemical approaches, the most important criterion is the number of calls to the calculation method, i.e., a stage limiting the performance of the entire approach. To determine the best approach, we rank the investigated methods by the average number of calls to the xTB calculation. The results are shown in Fig. 3. The best of the considered methods in terms of call numbers is the Artificial Bee Swarm (ABC) method, which is also one of the fastest approaches.

Thus, the ABC algorithm is the most promising of the global optimization evolutionary algorithms. In addition, the ABC algorithm is chosen for the following examination because of our desire not only to optimize the geometry, but also to find other low-energy possible conformations; the ABC algorithm allows one to get more data about the PES shape with a slight increase in computational resources.

We also change the initial bee population as number of DoFs multiplied by an integer multiplier instead of the original value equal to 10. On the one hand, it results in increasing the execution time of the first step, being able to flexibly vary the resources needed to optimize molecules of different sizes. On the other hand, it allows us to make a broader assessment of PES at the first optimization stage, which shifted the balance from exploitation to exploration search strategy. As a result, setting the population size equal to the number of degrees of freedom makes it possible to significantly reduce both the RMSD/DOF values and the number of required calls to the quantum chemical calculator. We test the original version of the ABC algorithm with the initial bee population as the number of DoFs multiplied by 10, the "ABC 1×DoF" approach with the coefficient equal to 1 instead of 10 and the "ABC 2×DoF" approach where the multiplier was equal to two (Fig. 4). The best mean value of RMSD/DoF, the energy difference between theoretical and experimental global minima ($\Delta$E/DoF), and the number of xTB calls are obtained for the ABC$\textunderscore$new$\textunderscore$1 approach, so next we use this implementation of the ABC method and the abbreviation ABC to designate the selected option.

\begin{figure}[t!]
  \centering
  \includegraphics[width=9cm]{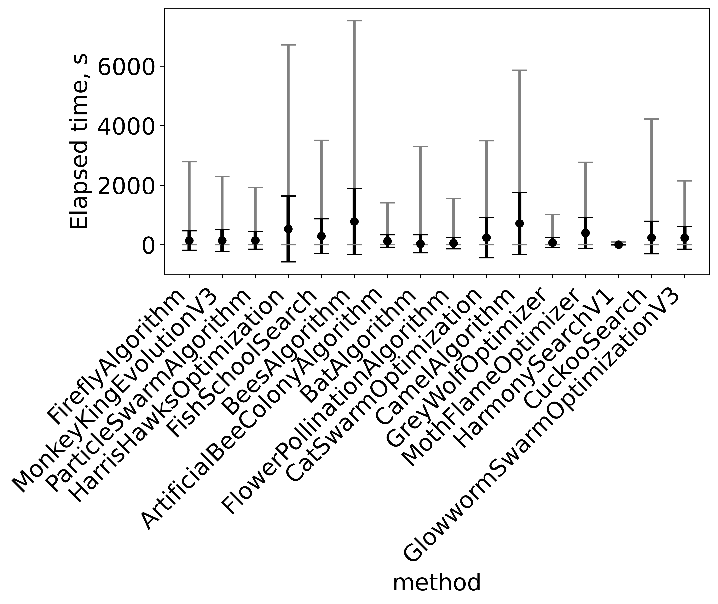}
  \caption{The average (point), 95\% confidence interval (black), and minimum and maximum (grey) time spent by various methods to find the equilibrium geometry of molecules.}
  \label{fig:fig2}
\end{figure}

Thus, we have shown that the ABC algorithm allows to successfully find the conformation of molecules corresponding to the global minima on PES. At the same time, this approach less often refers to the call of a quantum chemical program for calculating energy in comparison with many other methods of swarm intelligence. The choice of the initial population of bees equal to the number of degrees of freedom shows the best results in terms of consistency with experimentally obtained values.

However, a detailed consideration of PES is often of greater interest to scientists compared to a simple search for a global minimum. Due to the presence of molecules in a substance in different conformations, it is important to consider the existence of different conformers to accurately reproduce certain physicochemical and electronic properties of the system. Even though the electron density distribution strongly depends on the conformation of the molecule, today there is no single method for local conformers determination. Thus, our next step is to test the possibility of using the ABC algorithm to search for conformers, corresponding to local energy minima of organic molecules.

\subsection{Conformational analysis}
The ABC algorithm appeared to be the best method in terms of the number of calls of quantum chemical program. The algorithm is based on the implementation of the functions of three types of bees engaged in collecting nectar: employed bees—carry nectar from food sources and look around (determination of the known minimum position and local search in their vicinity), onlookers—watch the employed bees and direct them to new places (probabilistic assessment of the objective function shape), and scouts—look for new places if the employed bees do not succeed (if the minimum is not achieved in several consecutive moves, they randomly search for a new one).

The algorithm includes three parameters that could be varied: the number of bees in the initial population, number of the patience algorithm cycles, and fitness function, according to which the probability of choosing a given bee is estimated. The number of bees is defined initially as the number of DoFs, considering the accuracy and the operating speed discussed in the previous part. The number of cycles is set to five to maintain the balance between the optimization time and the time to search for a new conformation.

The role of the fitness function is to provide the widest possible exploration of PES. While the first step is the global minimum finding, the following stage is the search for the local ones. Since the conformations should be diverse, it is important to avoid sampling conformers in the same local minimum. Therefore, we need to get a new structure with geometry as different as possible relative to those already found. At the same time, we must try to reduce the energy of the system to converge to a new minimum. We use the sigmoid function as the basis of the fitness function to impose a penalty on the work of the bees when there is an increase in the geometric similarity of the new structure with the existing ones. We estimate the similarity of molecules by the RMSE calculation of the dihedral angles’ deviation of the new conformation with the structures already discovered. We also use the xTB-calculated energy as a multiplier to force the bees to search directly for energy minima, rather than arbitrary points on the PES. The final fitness function looks as follows:

\begin{equation}
  f=|{E}_{xtb}| \times \frac{1}{1+{10}^{5}\times exp(-10 \times {RMSE}_{best} )}
\end{equation}

where ${E}_{xtb}$ is the total energy of molecule calculated by the xTB, ${RMSE}_{best}$ is the minimum RMSD between the candidate conformation and already selected ones; the bees try to maximize the function. We also use the GFN2-xTB to prevent the case of the critical convergence of atoms.

\begin{figure}[t!]
  \centering
  \includegraphics[width=9cm]{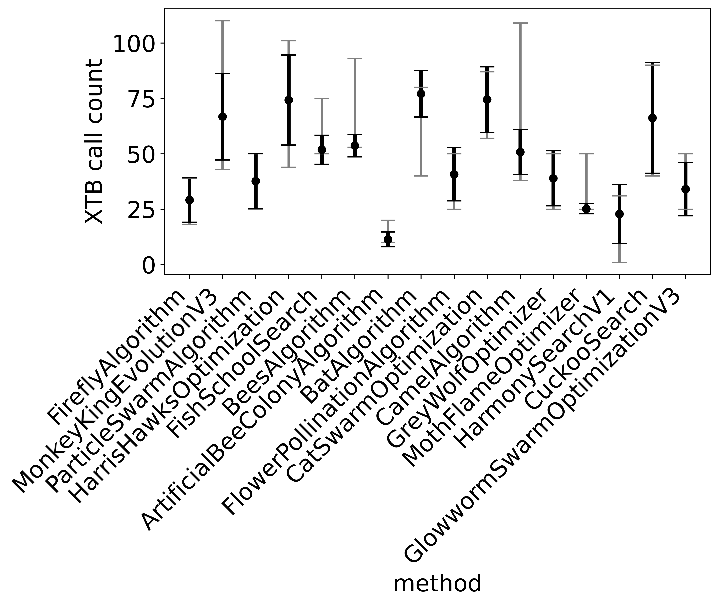}
  \caption{The average (point), 95\% confidence interval (black), and minimum and maximum (grey) number of calls to the quantum-chemical program for calculating the internal energy of given geometry.}
  \label{fig:fig3}
\end{figure}

We start with the gminorgDB\cite{mitrofanov2021iii} data analysis by the ABC approach with the parameters discussed. There were 11 organic molecules in the database with identified global and local minima on PES in a gas phase. The structures of the molecules are presented in Fig. 5. We use the ABC approach to find the global and local energy minima for each molecule in the dataset. Next, we compare the geometry similarity of the theoretical and experimental structures by calculation of the RMSD of non-hydrogen atomic positions. Since the RMSD of the atomic coordinates is a standard approach of the molecule geometries comparison in a literature, we present the results of the RMSD of the atomic positions instead of the RMSE of the dihedral angles, which we use in the fitness function.  We also compare the energies of the geometries by subtraction of the theoretical geometry energy value from the experimental one.

The detailed results for the molecule (1), hexan-2-one, as an example, are presented in Fig. 6. The same figures for all structures are added to the Supplementary Information (Fig. S2–S4). The molecule (1) has five degrees of freedom. The matrix of RMSD divided by DoF is shown in Fig. 6. The value of the RMSD of the atomic position divided by the number of the degree of freedom is 0.40 Å for the experimental and theoretical global minima. The best value of this parameter was obtained for the experimental global conformation and one of the theoretical local structures (loc 2 in Fig. 6); the value is equal to 0.31 Å. The molecule (1) has three local conformations in the database, and the ABC algorithm successfully identify all of them. We also check whether the geometries found belong to different minima by the Nudged Elastic Band (NEB) method implemented in the xTB. The graphs obtained has information about the behavior of the energy function along the reaction path from one conformer to another. The results are shown in Fig. 6. In some cases, we observe bond breaks when analyzing some conformers obtained within NEB; therefore, no reaction paths were presented. The bond break indicates the existence of a high potential barrier between the two energy minima. Nevertheless, for the molecule (1) we clearly create six reaction paths between different conformers and validated the presence of six different minima on the PES.

\begin{figure}[t!]
  \centering
  \includegraphics[width=15.5cm]{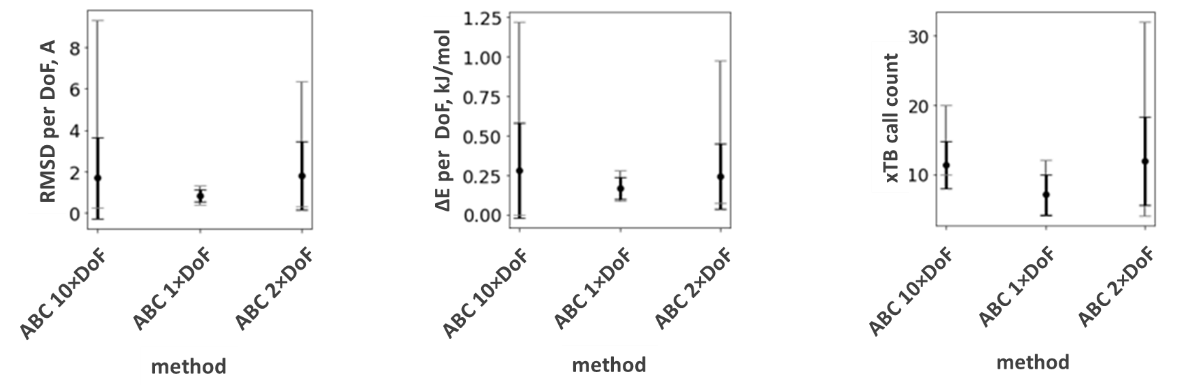}
  \caption{The average (point), 95\% confidence interval (black), and minimum and maximum (grey) resulting values of RMDS per DoF, $\Delta$E per DoF, and the number of calls to the calculation method for the original ABC algorithm (ABC 10×DoF) and two modified two modified versions of the algorithm (ABC 1×DoF and ABC 2×DoF).}
  \label{fig:fig4}
\end{figure}

The molecule (2) has nine local conformers but the value of RMSD/DOF between them shows that they have very similar geometry. The ABC algorithm identifies only one local conformer, which is very close to the experimental local ones. The RMSD/DOF value for the global conformations was 0.5 Å. The molecule (3) has one global and two local conformers in the database and the ABC approach discover them as well. The ABC algorithm detects three local geometries for the molecules (4)–(8), while the experimental database contains only two of them for each structure. The molecule (9) had two local confrontations, but the ABC approach does not find them, which may be due to its insensitivity to the conjugation between the double bond and carbonyl oxygen. The molecule (10) has three local conformers and we successfully discover them. While the ABC algorithm find one local structure for the last molecule (11), the experimental data has two of them. Nonetheless, we can suggest that the global and one of the experimental local structures have the same geometry because the value of RMSD/DoF between them is equal to zero.

\begin{table}[ht]
\centering
\caption{\label{tab:Table1}The values of RMSD/DoF and $\Delta$E obtained for the gas phase dataset with local and global structures.}
\setlength{\tabcolsep}{0.5em}
\renewcommand{\arraystretch}{1.4}
\begin{tabular}{ c c c c c c c }
 \hline
 \makecell{Molecule} & ${N}_{atoms}$	& DoF & RMSD/DoF, Å & E, kJ/mol & abs(E), kJ/mol & RMSD/DoF best, Å \\
 \hline
 1 & 7 & 5 & 0.40 & 0.07 & 0.07 & 0.31\\
 2 & 5 & 3 & 0.50 & –2.09 & 2.09 & 0.50\\
 3 & 9 & 1 & 1.50 & –0.25 & 0.25 & 1.33\\
 4 & 9 & 2 & 0.91 & 4.18 & 4.18 & 0.31\\
 5 & 7 & 1 & 0.86 & –13.62 & 13.62 & 0.03\\
 6 & 9 & 1 & 2.20 & 8.36 & 8.36 & 2.08\\
 7 & 12 & 3 & 0.90 & 16.72 & 16.72 & 0.87\\
 8 & 11 & 2 & 0.48 & 8.36 & –8.36 & 0.09\\
 9 & 5 & 2 & 0.45 & –4.42 & 4.42 & 0.34\\
 10 & 12 & 1 & 0.30 & 41.25 & 41.25 & 0.76\\
 11 & 9 & 2 & 0.98 & –16.33 & 16.33 & 0.98\\
 \hline
 & & mean & 0.86 & 2.32 & 10.51 & 0.69\\
\end{tabular}
\label{tab:tab1}
\end{table}

Generalized information about gas phase dataset with local and global structures is presented in Table 1. Here we show the number of atoms, DoF, RMSD/DoF values for the global experimental and theoretical conformations, the energy difference between them and RMSD/DoF values for the global experimental conformation, and the local one with the best similarity. The following mean values are obtained for the global conformers from the dataset and the ABC approach: the RMSD/DoF is 0.86 Å; $\Delta$E (for absolute values) is 10.51 kJ/mol. The energy error is similar to the xTB error. The RMSD/DoF values for the experimental global and the most similar theoretical one is 0.86 Å. We also compare the experimental global structure with the best theoretical local one as RMSD/DoF (best); the value is 0.69 Å.

The best similarity in the energy and the RMSD/DoF values are observed for molecules (1) and (2), which are linear molecules. The largest energy error is obtained for molecule (10), which has two heterocyclic aliphatic fragments in the structure. This fact could be explained as follows. Since most atoms are connected in cycles, even a small deviation of the values of one of the dihedral angles in the system leads to a displacement of all atoms and a sharp increase in the RMSD/DoF value. We also observe a large deviation in energy in the case of another heterocyclic molecule (5). There are also relatively large errors in a case of the (7) and (11) molecules due to the error of strongly conjugated systems. The achieved results, however, indicate the successful identification of global and local conformations, thereby providing a rationale for applying the algorithm to crystalline structures.

\subsection{Organic crystal packaging}
Since the theoretical definition of the molecule conformation in a gas phase could be realized in different ways (as a geometry optimization, conformation generation by various toolkits as RDKit, etc.), the question about the crystal structure prediction does not have a solution. There are two main factors will influence in bulk molecular crystal case: the number of stable conformations and, for a given conformation, the effect of packing forces on the molecular geometry, because the equilibrium geometry of the molecule in a crystal is a balance of intramolecular and intermolecular forces. It is worth noting that the two aforementioned influences frequently operate at varying magnitudes. Specifically, the intermolecular interactions are notably weaker than covalent bonds responsible for molecular packaging. Thus, it is expected that the energetically preferable geometry of the isolated molecule is generally a good starting point for generating possible crystal structures. Therefore, our next step is to validate the ABC approach for the solid-state dataset.

\begin{figure}[t!]
  \centering
  \includegraphics[width=12cm]{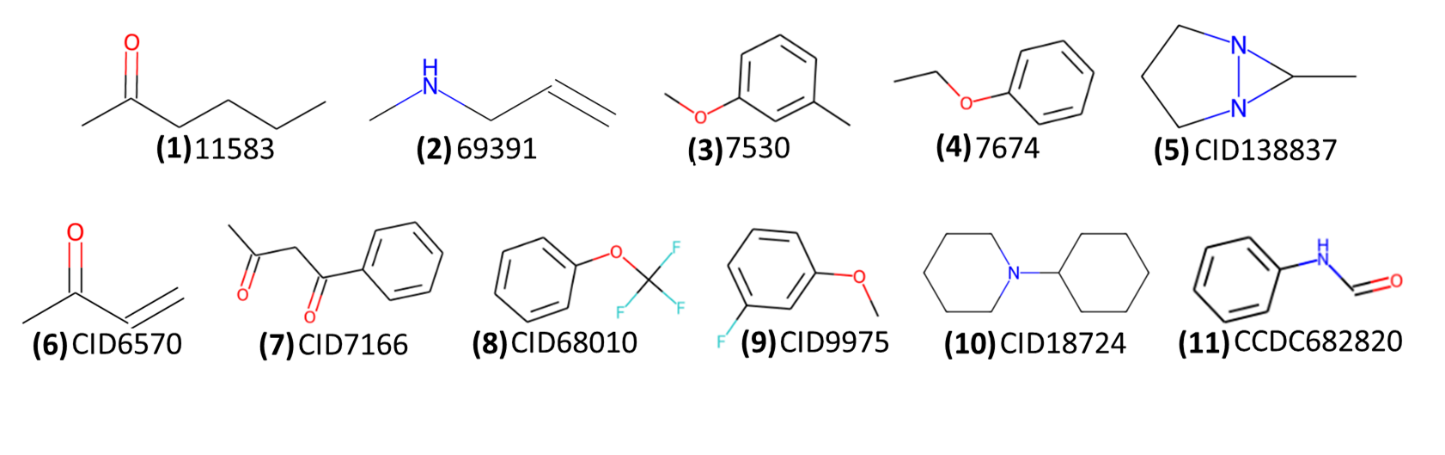}
  \caption{Structures of molecules from the gas phase dataset with the data about local and global energy minima conformations. The bold font number was used in the text of the article, the thin font number corresponded to the gminorgDB names.}
  \label{fig:fig5}
\end{figure}

We also use the experimental data from the crystal dataset\cite{nangia2008conformational}. It consists of 20 different molecules with a number of conformations; all structures are presented in Fig. 7. The detailed results of our study are added to SI (Fig. S5–S7) and Table 2. Since there is no concept for finding a global energy minimum for the molecular conformation in the periodic crystal structure, we compare the results for RMSD/DoF for the best matches between the experimental and theoretical data. The $\Delta$E value is calculated for the conformations with the best RMSD/DoF. The obtained results for the RMSD/DoF are very close to the gas phase and we can declare that the ABC approach could successfully find the conformations of molecules in the solid state. However, the energy difference between conformers are much higher for the crystal case. The mean value was –360.96 kJ/mol (instead of the 2.32 kJ obtained in the gas-phase), and the energies of the theoretical conformers are always higher than the energy of the experimental ones. This fact indicates the stabilization of the structure of the crystal field, and the average value is similar to $\Delta$E. Meanwhile the solid phase conformation is not in the local or global minima because of crystal field stabilization, we optimize the geometry of the best crystal conformation in terms of the RMSD/DoF value by the xTB approach. Next, we compare the energy of optimized conformation energy with the theoretical suggested by the ABC. Results are presented in Table 2.

\begin{figure}[t!]
  \centering
  \includegraphics[width=16cm]{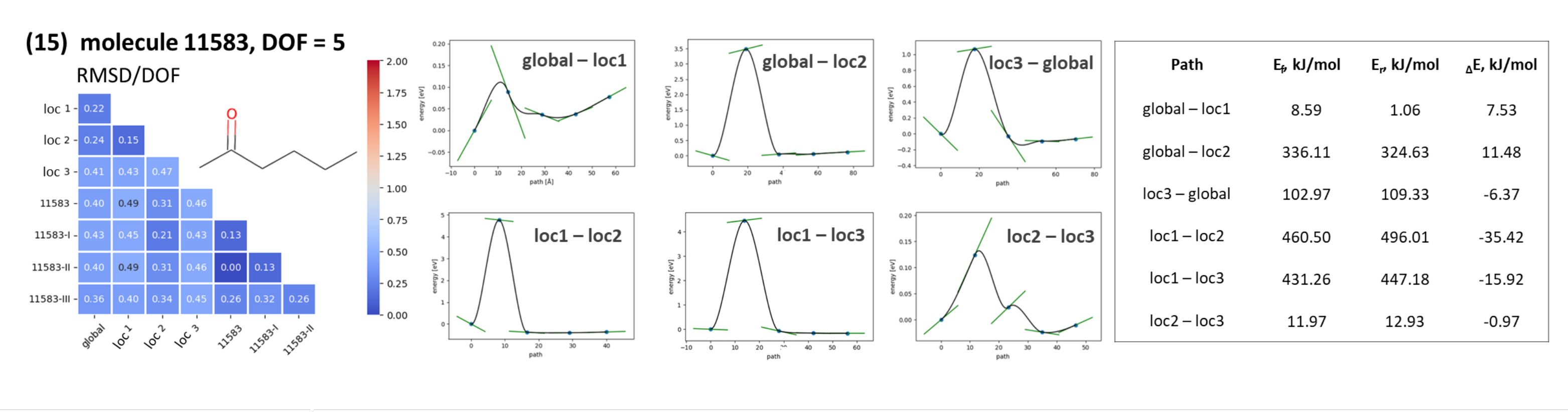}
  \caption{The detailed analysis of the theoretical and experimental conformations of the molecule (1). The ${E}_{f}$ value was equal to the activation energy, ${E}_{r}$ was the difference between the lowest and the highest values of energy during the path and $\Delta$E was the energy difference for the conformers analyzed.}
  \label{fig:fig6}
\end{figure}

\begin{figure}[t!]
  \centering
  \includegraphics[width=16cm]{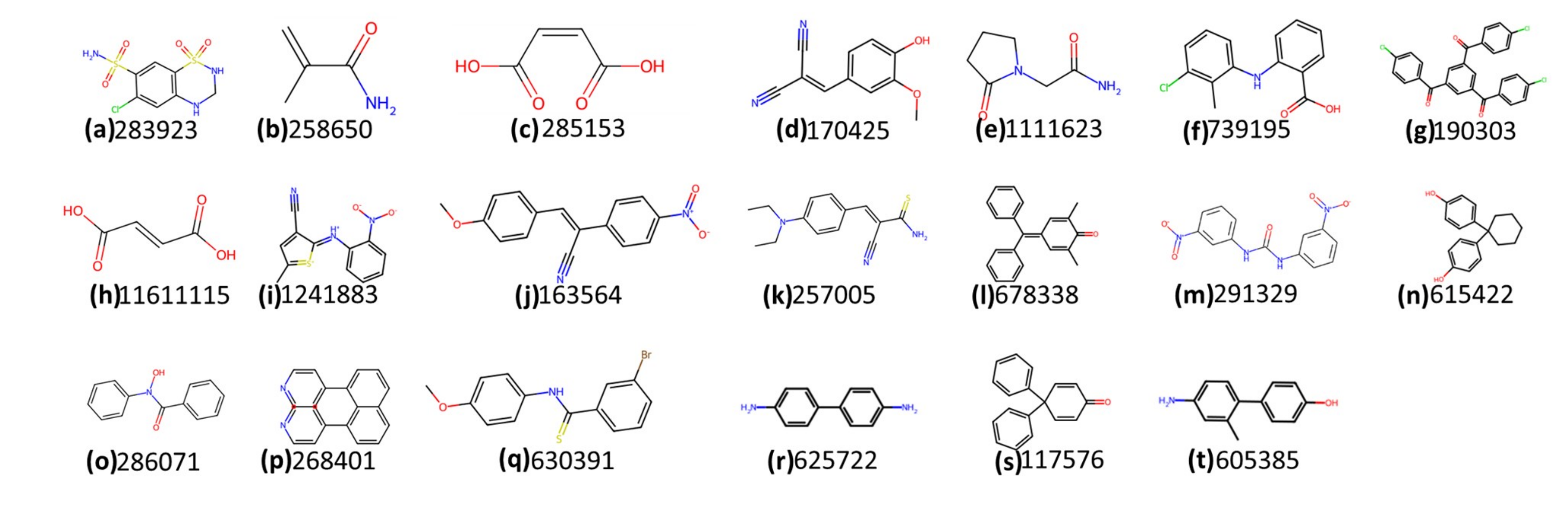}
  \caption{Structures of molecules from the crystal phase dataset. The bold font number is used in the text of the article, the thin font number correspond to the ID numbers of phases from the Cambridge Structural Database (CCDC).}
  \label{fig:fig7}
\end{figure}

We also compare the ABC algorithm with the Fast Small-Molecule Conformer Generators approach suggested by Wang and colleagues in the recent study\cite{wang2023small}. Since their approach has been developed using the crystal data, we generate the conformers for (a)–(t) molecules. The number of conformers is determined according to the original recommendations. The algorithm fails to process six molecules ((e), (f), (m)–(p)). The mean value of RMSD/DoF values is 1.01 Å for the successfully calculated molecules. The detailed results for those molecules that the approach processed are presented in the Supplementary Information (Fig. S8–S10). Thereby the ABC approach developed in the present paper allowed to reduce the error in the conformation determination in two times. Thus, we hope that our approach allows community to get closer to the solution of the organic crystal packaging problem.

\begin{table}[ht]
\centering
\caption{\label{tab:Table2}The values of RMSD/DoF and $\Delta$E obtained for the crystal phase molecules dataset.}
\setlength{\tabcolsep}{0.5em}
\renewcommand{\arraystretch}{1.4}
\begin{tabular}{ c c c c c c c }
 \hline
 \makecell{Molecule} & ${N}_{atoms}$	& DoF & RMSD/DoF (best), Å & $\Delta$E, kJ/mol (best) & $\Delta$E, kJ/mol (opt)\\
 \hline
 a & 17 & 1 & 2.17 & –225.61 & 1.93\\
 b & 6 & 1 & 1.42 & –127.66 & 10.56\\
 c & 8 & 2 & 0.03 & –88.28 & 0.60\\
 d & 13 & 2 & 0.04 & –331.39 & 18.92\\
 e & 10 & 2 & 0.42 & –353.59 & 6.45\\
 f & 18 & 3 & 0.37 & –531.70 & 54.60\\
 g & 33 & 6 & 0.39 & –583.95 & 9.04\\
 h & 8 & 2 & 0.04 & –35.86 & 20.82\\
 i & 18 & 3 & 0.79 & –336.45 & 4.78\\
 j & 21 & 4 & 0.12 & –466.49 & 5.11\\
 k & 18 & 5 & 0.26 & –321.44 & 3.66\\
 l & 22 & 2 & 1.03 & –353.59 & 13.22\\
 m & 22 & 8 & 0.35 & –495.23 & 11.45\\
 n & 20 & 2 & 0.46 & –113.37 & 33.75\\
 n & 16 & 3 & 0.55 & –432.91 & 12.21\\
 p & 22 & 2 & 0.42 & –514.14 & 10.27\\
 q & 18 & 4 & 0.28 & –476.52 & 2.35\\
 r & 14 & 1 & 0.11 & –393.17 & 8.16\\
 s & 19 & 2 & 0.07 & –513.30 & 3.24\\
 t & 15 & 1 & 1.63 & –524.59 & 4.21\\
 \hline
 & & mean & 0.55 & –360.96 & 11.77\\
\end{tabular}
\label{tab:tab2}
\end{table}

\section{Conclusion}
Our calculations show that the bee swarm intelligence method and the intramolecular parameters such as torsion (dihedral) angles can be effectively used to predict both isolated conformers in gas phase and conformation of molecules in periodic molecular crystals. The fitness function proposed in the study allows one to effectively search local conformers, considering the wide possible variation of conformer geometries considering the RMSD value and the purpose to find the energy minimum. Despite the limited set of verification organic molecule dataset, we hope that the developed approach will be useful for scientists working both with gas-phase reactions and studying the chemistry of solutions. In addition, we hope that our work will be an important step towards solving open question of the packaging of organic molecules in crystals.

\section{Acknowledgements}
\label{sec:acknowledgements}
The investigations were supported by financing Interdisciplinary Scientific and Educational Schools of Moscow M.V. Lomonosov State University (No. 23-Ш03-04). The research was carried out using the equipment of the shared research facilities of HPC computing resources at Lomonosov Moscow State University.

\bibliographystyle{unsrt}
\bibliography{references}

\begin{thebibliography}{10}

\bibitem{mitrofanov2021iii}
Artem Mitrofanov, Nikolai Andreadi, Petr Matveev, Gladis Zakirova, Nataliya Borisova, Stepan Kalmykov, and Vladimir Petrov.
\newblock An (iii)/ln (iii) solvent extraction: Theoretical and experimental investigation of the role of ligand conformational mobility.
\newblock {\em Journal of Molecular Liquids}, 325:115098, 2021.

\bibitem{rankin2013structural}
David~WH Rankin, Norbert Mitzel, and Carole Morrison.
\newblock {\em Structural methods in molecular inorganic chemistry}.
\newblock John Wiley \& Sons, 2013.

\bibitem{townes2013microwave}
Charles~H Townes and Arthur~L Schawlow.
\newblock {\em Microwave spectroscopy}.
\newblock Courier Corporation, 2013.

\bibitem{broyden1967quasi}
Charles~G Broyden.
\newblock Quasi-newton methods and their application to function minimisation.
\newblock {\em Mathematics of Computation}, 21(99):368--381, 1967.

\bibitem{thompson2014conformations}
Hugh~PG Thompson and Graeme~M Day.
\newblock Which conformations make stable crystal structures? mapping crystalline molecular geometries to the conformational energy landscape.
\newblock {\em Chemical Science}, 5(8):3173--3182, 2014.

\bibitem{fletcher1970new}
Roger Fletcher.
\newblock A new approach to variable metric algorithms.
\newblock {\em The computer journal}, 13(3):317--322, 1970.

\bibitem{goldfarb1970family}
Donald Goldfarb.
\newblock A family of variable-metric methods derived by variational means.
\newblock {\em Mathematics of computation}, 24(109):23--26, 1970.

\bibitem{shanno1970conditioning}
David~F Shanno.
\newblock Conditioning of quasi-newton methods for function minimization.
\newblock {\em Mathematics of computation}, 24(111):647--656, 1970.

\bibitem{bitzek2006structural}
Erik Bitzek, Pekka Koskinen, Franz G{\"a}hler, Michael Moseler, and Peter Gumbsch.
\newblock Structural relaxation made simple.
\newblock {\em Physical review letters}, 97(17):170201, 2006.

\bibitem{goedecker2004minima}
Stefan Goedecker.
\newblock Minima hopping: An efficient search method for the global minimum of the potential energy surface of complex molecular systems.
\newblock {\em The Journal of chemical physics}, 120(21):9911--9917, 2004.

\bibitem{andreadi2022tree}
Nikolai Andreadi, Dmitry Zankov, Kirill Karpov, and Artem Mitrofanov.
\newblock Tree parzen estimator for global geometry optimization: A benchmark and database of experimental gas-phase structures of organic molecules.
\newblock {\em Journal of Computational Chemistry}, 43(21):1434--1441, 2022.

\bibitem{riniker2015better}
Sereina Riniker and Gregory~A Landrum.
\newblock Better informed distance geometry: using what we know to improve conformation generation.
\newblock {\em Journal of chemical information and modeling}, 55(12):2562--2574, 2015.

\bibitem{wang2023small}
Zhe Wang, Haiyang Zhong, Jintu Zhang, Peichen Pan, Dong Wang, Huanxiang Liu, Xiaojun Yao, Tingjun Hou, and Yu~Kang.
\newblock Small-molecule conformer generators: Evaluation of traditional methods and ai models on high-quality data sets.
\newblock {\em Journal of Chemical Information and Modeling}, 63(21):6525--6536, 2023.

\bibitem{yang2010firefly}
Xin-She Yang.
\newblock Firefly algorithm, stochastic test functions and design optimisation.
\newblock {\em International journal of bio-inspired computation}, 2(2):78--84, 2010.

\bibitem{eberhart1995new}
Russell Eberhart and James Kennedy.
\newblock A new optimizer using particle swarm theory.
\newblock In {\em MHS'95. Proceedings of the sixth international symposium on micro machine and human science}, pages 39--43. Ieee, 1995.

\bibitem{pham2006bees}
Duc~Truong Pham, Afshin Ghanbarzadeh, Ebubekir Ko{\c{c}}, Sameh Otri, Shafqat Rahim, and Muhamad Zaidi.
\newblock The bees algorithm—a novel tool for complex optimisation problems.
\newblock In {\em Intelligent production machines and systems}, pages 454--459. Elsevier, 2006.

\bibitem{karaboga2005idea}
Dervis Karaboga et~al.
\newblock An idea based on honey bee swarm for numerical optimization.
\newblock Technical report, Technical report-tr06, Erciyes university, engineering faculty, computer~…, 2005.

\bibitem{gonzalez2008nature}
Juan~R Gonz{\'a}lez, Alejandro Sancho-Royo, David~A Pelta, and Carlos Cruz.
\newblock Nature-inspired cooperative strategies for optimization.
\newblock In {\em Encyclopedia of Networked and Virtual Organizations}, pages 982--989. IGI Global, 2008.

\bibitem{yang2012flower}
Xin-She Yang.
\newblock Flower pollination algorithm for global optimization.
\newblock In {\em International conference on unconventional computing and natural computation}, pages 240--249. Springer, 2012.

\bibitem{ibrahim2016novel}
Mohammed~Khalid Ibrahim and Ramzy~Salim Ali.
\newblock Novel optimization algorithm inspired by camel traveling behavior.
\newblock {\em Iraqi Journal for Electrical and Electronic Engineering}, 12(2):167--177, 2016.

\bibitem{heidari2019harris}
Ali~Asghar Heidari, Seyedali Mirjalili, Hossam Faris, Ibrahim Aljarah, Majdi Mafarja, and Huiling Chen.
\newblock Harris hawks optimization: Algorithm and applications.
\newblock {\em Future generation computer systems}, 97:849--872, 2019.

\bibitem{chu2006cat}
Shu-Chuan Chu, Pei-Wei Tsai, and Jeng-Shyang Pan.
\newblock Cat swarm optimization.
\newblock In {\em PRICAI 2006: Trends in Artificial Intelligence: 9th Pacific Rim International Conference on Artificial Intelligence Guilin, China, August 7-11, 2006 Proceedings 9}, pages 854--858. Springer, 2006.

\bibitem{mirjalili2014grey}
Seyedali Mirjalili, Seyed~Mohammad Mirjalili, and Andrew Lewis.
\newblock Grey wolf optimizer.
\newblock {\em Advances in engineering software}, 69:46--61, 2014.

\bibitem{shehab2020moth}
Mohammad Shehab, Laith Abualigah, Husam Al~Hamad, Hamzeh Alabool, Mohammad Alshinwan, and Ahmad~M Khasawneh.
\newblock Moth--flame optimization algorithm: variants and applications.
\newblock {\em Neural Computing and Applications}, 32:9859--9884, 2020.

\bibitem{krishnanand2009glowworm}
KN~Krishnanand and Debasish Ghose.
\newblock Glowworm swarm optimisation: a new method for optimising multi-modal functions.
\newblock {\em International Journal of Computational Intelligence Studies}, 1(1):93--119, 2009.

\bibitem{gandomi2013cuckoo}
Amir~Hossein Gandomi, Xin-She Yang, and Amir~Hossein Alavi.
\newblock Cuckoo search algorithm: a metaheuristic approach to solve structural optimization problems.
\newblock {\em Engineering with computers}, 29:17--35, 2013.

\bibitem{bastos2008novel}
Carmelo~JA Bastos~Filho, Fernando~B de~Lima~Neto, Anthony~JCC Lins, Antonio~IS Nascimento, and Marilia~P Lima.
\newblock A novel search algorithm based on fish school behavior.
\newblock In {\em 2008 IEEE international conference on systems, man and cybernetics}, pages 2646--2651. IEEE, 2008.

\bibitem{geem2001new}
Zong~Woo Geem, Joong~Hoon Kim, and Gobichettipalayam~Vasudevan Loganathan.
\newblock A new heuristic optimization algorithm: harmony search.
\newblock {\em simulation}, 76(2):60--68, 2001.

\bibitem{meng2016monkey}
Zhenyu Meng and Jeng-Shyang Pan.
\newblock Monkey king evolution: a new memetic evolutionary algorithm and its application in vehicle fuel consumption optimization.
\newblock {\em Knowledge-Based Systems}, 97:144--157, 2016.

\bibitem{vrbanvcivc2018niapy}
Grega Vrban{\v{c}}i{\v{c}}, Lucija Brezo{\v{c}}nik, Uro{\v{s}} Mlakar, Du{\v{s}}an Fister, and Iztok Fister.
\newblock Niapy: Python microframework for building nature-inspired algorithms.
\newblock {\em Journal of Open Source Software}, 3(23):613, 2018.

\bibitem{o2011open}
Noel~M O'Boyle, Michael Banck, Craig~A James, Chris Morley, Tim Vandermeersch, and Geoffrey~R Hutchison.
\newblock Open babel: An open chemical toolbox.
\newblock {\em Journal of cheminformatics}, 3(1):1--14, 2011.

\bibitem{bannwarth2021extended}
Christoph Bannwarth, Eike Caldeweyher, Sebastian Ehlert, Andreas Hansen, Philipp Pracht, Jakob Seibert, Sebastian Spicher, and Stefan Grimme.
\newblock Extended tight-binding quantum chemistry methods.
\newblock {\em Wiley Interdisciplinary Reviews: Computational Molecular Science}, 11(2):e1493, 2021.

\bibitem{henkelman2000climbing}
Graeme Henkelman, Blas~P Uberuaga, and Hannes J{\'o}nsson.
\newblock A climbing image nudged elastic band method for finding saddle points and minimum energy paths.
\newblock {\em The Journal of chemical physics}, 113(22):9901--9904, 2000.

\bibitem{nangia2008conformational}
Ashwini Nangia.
\newblock Conformational polymorphism in organic crystals.
\newblock {\em Accounts of chemical research}, 41(5):595--604, 2008.

\end{thebibliography}

\end{document}